\documentstyle[preprint,aps]{revtex}
\input epsf.tex
\def\DESepsf(#1 width #2){\epsfxsize=#2 \epsfbox{#1}}
\begin{document}
\preprint{\vbox{\hbox{}}}
\draft
\title{
Extra Dimensions and Higgs Pair Production at Photon Colliders}
\author{Xiao-Gang He\footnote{E-mail: hexg@phys.ntu.edu.tw}}
\address{
Department of Physics, National Taiwan University, Taipei, Taiwan 10617, R.O.C.
}
\date{May 1999}
\maketitle
\begin{abstract}
We show that new physics effects due to extra dimensions can dramatically affect
Higgs pair production at photon colliders.
We find that the cross section due to extra dimensions with the scale $M_S$ of new
physics around $1.5$ TeV, the cross section
can be as large as 0.11 pb (1.5pb) for monochromatic photon collision,
$\gamma \gamma \to HH$, with
the collider energy $\sqrt{s} = 0.5 (1) $ TeV for Higgs mass of 100 (350) GeV.
The cross section can be 3 fb (2.7 fb) for the same parameters for collisions
using photon beams from electron or positron back scattered by laser.
These cross sections are
much larger than those predicted in the Standard Model. Higgs pair production at photon
colliders can provide
useful tests for new physics due to extra dimensions.

\end{abstract}
\newpage
It has recently been proposed
that gravitational effects can become large at a scale
$M_S$ near the weak scale due to effects from extra dimensions\cite{1,1a},
quite different from the traditional thought that gravitational effects
only become large at the Planck scale $M_{Pl} \sim 10^{19}$ GeV.
In this proposal the total space-time is $D=4 + n$ dimensions.
When the extra dimensions are compactified
there are towers of
states, the Kaluza-Klein (KK) states, interact with ordinary matter fields.
Although the interactions with the Standard Model (SM) fields for each of the KK state
is small, proportional to the Newton constant $G_N$,
the effects become much stronger, proportional to $1/M_S^2$, when the contributions
of all the KK states are summed
over.

The relation between the scale $M_S$ and the Planck scale
$M_{Pl}$, assuming all extra dimensions are compactified with the
same size R, is given by, $M^2_{Pl} \sim R^n M^{2+n}_S$. With
$M_S$ near a TeV, for n= 1 $R$ would be too large which 
is ruled out. However, with n larger than or equal to
2, the theory is not ruled out. In these cases, $R$ can be in the
sub-millimeter region which can be probed by laboratory
experiments\cite{2,3,4,5,6,6a}. The lower bound for $M_S$ is
constrained, typically, to be of order one TeV from present
experimental data\cite{2,3,4,5,6,6a}. 
Future experiments will provide more stringent
constraints. It is important to investigate as many systems as
possible to give constraints on the scale $M_S$ and to look for
deviations from predictions in the to isolate effects due to
extra dimensions. In this paper we study
effects due to extra dimensions 
on Higgs pair production at photon colliders. We find that
the cross section for this process can be dramatically different from 
SM predictions.

In the minimal Standard Model there is a neutral Higgs boson $H$
resulting from spontaneous symmetry breaking of $SU(2)_L\times
U(1)_Y$ to $U(1)_{em}$ due to the Higgs  mechanism. The mechanism
for spontaneous symmetry breaking is not well understood. There is
no experimental evidence favoring any particular mechanism, such
as the Higgs mechanism. The discovery of the Higgs boson and
understanding of its properties are fundamentally
important\cite{7}. The Higgs boson is also the last particle yet
to be discovered in the minimal SM. The discovery of Higgs boson
is one of the most important goals for future high energy
colliders. At present the lower bound on the SM Higgs boson mass
$m_h$ is set by LEP II to be 95.5 GeV at 95\% C.L.\cite{8}. Many
methods have been proposed to produce and to study the properties
of Higgs bosons\cite{7}. Higgs pair production at various
colliders are some of the interesting ones\cite{9,10,11}. 
In
particular Higgs pair production through $\gamma \gamma \to H H$
is important for the study of Higgs boson properties.
In SM the
process $\gamma \gamma \to HH$ only occur at loop level with cross
section of order O(0.5 fb) for $\sqrt{s} \sim (500\;\;
\mbox{to}\;\; 1000)$ and $m_h$ in the range 100 to 400
GeV\cite{10}. It is in principle reachable\cite{12}, but a
formidable challenge for accelerator physics. The smallness of
cross section predicted in SM makes the process, $\gamma \gamma
\to HH$, sensitive to new physics beyond the minimal SM. New
physics beyond SM may dramatically change the situation and
increase the cross section to a level more practical for
experiments\cite{11}. In the following we show that this is indeed
possible with extra dimensions.

After compactifying the extra n dimensions, for a given KK level
$\vec l$ there are one spin-2 state, n-1 spin-1 and n(n-1)/2
spin-0 states\cite{6}. Assuming that all SM fields are confined to
a four dimensional world-volume and gravitation is minimally
coupled to SM fields, it was found that the spin-1 KK states
decouple while the spin-2 and spin-0 KK states couple to all SM
fields\cite{6}. Since the KK states interact with particles in the
SM, exchanges of KK states can generate new interactions among the
SM particles. For $\gamma \gamma \to HH$ we find that contrary to
the situation in SM, this process can happen at the tree level due
to exchange of spin-2 KK states as shown in Fig.1. Spin-1 and
spin-0 KK states do not contribute to this process. Using the
Feynmann rules given in Ref.\cite{6}, we have

\begin{eqnarray}
&&\widetilde M(\gamma \gamma \to HH)
= {\kappa^2\over 8} (m^2_h\eta^{\mu\nu} -
C^{\mu\nu,\rho\sigma} k_{1\rho} k_{2\sigma})\nonumber\\
&&\;\;\;\;\;\;\;\;\;\;\times {B_{\mu\nu,\alpha\beta}\over q^2-m_l^2}
(p_1\cdot p_2 C^{\alpha\beta,\delta\gamma}
+ D^{\alpha\beta,\delta\gamma}) \epsilon_{1\delta}(p_1)
 \epsilon_{2\gamma}(p_2),\nonumber\\
&&C^{\mu\nu,\rho\sigma} = \eta^{\mu\rho}\eta^{\nu\sigma}
+\eta^{\mu\sigma}\eta^{\nu\rho}-\eta^{\mu\nu} \eta^{\rho\sigma},\nonumber\\
&&D^{\alpha\beta,\delta\gamma}
= \eta^{\alpha\beta} k_1^\delta k_2^\gamma
-[\eta^{\alpha\gamma} k_1^\beta k_2^\delta + \eta^{\alpha \delta}
k_1^\gamma k_2^\beta
-\eta^{\delta\gamma} k_1^\alpha k_2^\beta + (\alpha \to \beta,
\beta\to \alpha)],\nonumber\\
&&B_{\mu\nu,\alpha\beta}
= (\eta_{\mu\alpha}- {q_\mu q_\alpha\over m_l^2})
(\eta_{\nu\beta}-{q_\nu q_\beta
\over m_l^2})+(\eta_{\nu\alpha} - {q_\nu q_\alpha\over m^2_l})
(\eta_{\mu\beta}-{q_\mu q_\beta\over m_l^2})\nonumber\\
&&\;\;\;\;\;\;\;\;\;\;-{2\over 3} (\eta_{\mu\nu} -
{q_\mu q_\nu\over m_l^2})
(\eta_{\alpha\beta} - {q_\alpha q_\beta\over m_l^2}).
\end{eqnarray}
where $\kappa^2 = 16\pi G_N$, and $m_{l}$ is KK
state mass.
After a straight forward calculation and 
summing over all KK state contributions, we have
\begin{eqnarray}
M &=& -i{\kappa^2 D_n(s) \over 4}
[ 2 u \epsilon_1\cdot k_1
\epsilon_2\cdot k_2 + 2t \epsilon_1\cdot k_2 \epsilon_2\cdot k_1\nonumber\\
&+&\epsilon_1\cdot \epsilon_2 (t u -m_h^4) +2 m_h^2
( \epsilon_1\cdot k_1 \epsilon_2\cdot k_1
+ \epsilon_1 \cdot k_2 \epsilon_2\cdot k_2)],
\end{eqnarray}
where $s=q^2$, $t=(k_1-p_1)^2=(p_2-k_2)^2$,
$u=(k_1-p_2)^2=(k_2-p_1)^2$, $D_n(s) = \sum_l i/(s-m_l^2)$.
With the cut-off scale for the summation
set to be the same as $M_S$, one has

\begin{eqnarray}
&&\kappa^2 D_n(s) =
{16\pi \over M_S^4} \left ({s\over M_S^2} \right )^{n/2-1}
[\pi + i2 I_n(M_S/\sqrt{s})],
\nonumber\\
&&I_n(x)
=\left \{ \begin{array}{ll}
-\sum^{n/2-1}_{k=1} {1\over 2k} x^{2k} - {1\over 2} \ln(x^2-1),&
 \;\;\;\;\mbox{for}\;\;n=\mbox{even},\\
-\sum^{(n-1)/2}_{k=1} {1\over 2k-1}
x^{2k-1} + {1\over 2} \ln {x+1\over x-1},&\;\;\;\;\mbox{for}\;\;n=\mbox{odd}.
\end{array}
\right .
\end{eqnarray}
For $M_S^2 \gg s$,
$\kappa^2 D_n(s)$ is approximately equal to $i (16\pi/M_S^4) \ln (M_S^2/s)$ for
$n=2$ and $i(16\pi/M_S^4) 2/(n-2)$ for $n>2$. There are a
few different definitions of $M_S$ used in the literature, although the same order of
magnitude there are some differences numerically. We will use
$M_S$ defined above for our discussions\cite{6a}.

The helicity amplitudes for different photon polarizations are given by

\begin{eqnarray}
&&M((1,+),(2,+))=M((1,-),(2,-))={\kappa^2 D_n(s)\over 16}s^2
(1-{4m_h^2\over s})\sin^2\theta,
\nonumber\\
&&M((1,+),(2,-))=M((1,-),(1,+)) = 0,
\end{eqnarray}
where $\theta$ is the angle between the photon beam line and the out going Higgs boson
momentum. $(1,\pm)$ and $(2,\mp)$ indicate the photon polarization vectors with
$\epsilon_{1,\mu}(\pm)=\epsilon_{2,\mu} (\mp) = (0,\mp1,0,-i)/\sqrt{2}$.
The cross section for polarized photons is given by,

\begin{eqnarray}
{d\sigma ((1,\pm), (2,\pm))\over d\cos\theta}
&=& {\pi \over 64 M_S^2} \left ({s\over M_S^2} \right )^{n+1}
[\pi^2 + 4 I^2_n (M_S/\sqrt{s})]
\sin^4\theta \left ( 1- {4m_h^2\over s}\right )^{5/2},\nonumber\\
{d\sigma ((1,\pm), (2,\mp))\over d\cos\theta}&=&0.
\label{gg}
\end{eqnarray}
The cross section for the unpolarized photon beam, $d\sigma(s)/d\cos\theta$ 
is the same
as the above.
We will use this case for later discussions.

In SM the lowest order contributions to $\gamma\gamma \to HH$
occur at one loop level. The cross section is predicted to be
small. For monochromatic photon beam with $\sqrt{s} =0.5 (1)$ TeV
and the Higgs mass $m_h$ in the range of $100\sim 200 (100\sim
400)$ GeV, the cross section $\sigma(s)$ is in the range $0.8\sim
0.4 (0.6\sim 0.5)$ fb\cite{10}. These cross sections are small and
difficult to achieve experimentally although not impossible. With
effects from extra dimensions the cross section can be
dramatically different.

The effects of extra dimensions on the Higgs pair production at photon colliders depend
on the scale $M_S$, the number of extra dimensions n, the Higgs boson mass $m_h$ and
the collider energy $\sqrt{s}$. The cross section increases rapidly with $s$ which is different
than the prediction of SM where the cross section does not change dramatically with $\sqrt{s}$. The
cross section decreases when $M_S$, $m_h$ and n are increases.
In our numerical analyses, we will
take two typical value 1.0 and 1.5 TeV for the scale $M_S$
and vary
n up to 7 and $m_h$ in typical kinematic allowed ranges for illustration
\footnote{There is a singularity
when $\sqrt{s} = M_S$ due to our negligence of the decays
of the KK states. To avoid this artificial singularity, for $\sqrt{s} =1$ TeV
we use $M_S = 1.01$ TeV such that $M_S$ is
sufficiently
away from $\sqrt{s}$.}.

The results for $\sigma(s)$ with
monochromatic photon beams are shown in Table 1.
From Table 1, we see that for $\sqrt{s} = 0.5$ TeV and $m_h$ as high
as 200 GeV,
the cross sections can be as large as 14 fb and 274 fb for $M_S=1.5$ TeV and 1 TeV, respectively.
Even with n=7, the cross section can still reach 11 fb for $M_S = 1$ TeV.
For $\sqrt{s} = 1$ TeV, the cross sections are even larger.
For the worst case of n =7, for $m_h$ as large as 450 GeV the cross section can still reach 12 fb
and 3.9 pb with $M_S = 1.5$ TeV and 1.01 TeV, respectively.  We also analyzed the case
with $\sqrt{s} =250$ GeV. In this case, the cross section for
$M_S = 1.5$ TeV the cross section is small ($< 0.34$ fb) with $M_S = 1.5$ TeV. But
the cross section can still be as large as 6.6 fb for $m_h = 100$ GeV with $M_S=1.01$ TeV.
It is clear that the cross sections for Higgs pair production at photon colliders
with extra dimensions can be much larger than the
SM predictions.

High energy monochromatic photon beams may be difficult to obtain.
A practical method to obtain high energy photon beams is to use
laser back-scattering technique on an electron or positron beam
which produces abundant numbers of hard photons nearly along the
same direction as the original electron or positron beam. The
photon beam energy obtained this way is not monochromatic. The
energy spectrum of the back-scattered photon is given by\cite{12}

\begin{eqnarray}
&&F(x) = {1\over D(\xi)} [1-x + {1\over 1-x}-{4x\over \xi (1-x)}
+{4x^2\over \xi^2(1-x)^2}],\nonumber\\
&&D(\xi) = (1-{4\over \xi} - {8\over \xi^2})\ln (1+\xi) + {1\over 2} +{8\over \xi}
-{1\over 2(1+\xi)^2},
\end{eqnarray}
where $x$ is the fraction of the energy of the incident electron. The parameter
$\xi$ is determined to be $2(1+\sqrt{2})$ by requiring that the back-scattered photon to
have the largest possible energy, but
does not interfere with the incident photon to create unwanted $e^+ e^-$. The maximal
value $x_{max}$ can be reach by x is given by $\xi/(1+\xi)\approx 0.828$.

The cross section $\sigma(e^+e^-)$ due to the back-scattered photon beams collide to
produce Higgs pair, $e^+ e^- \to \gamma \gamma \to HH$, is obtained by folding in the photon
luminosities with the cross section $\sigma(s)$. We have

\begin{eqnarray}
\sigma(e^+e^-)
=\int^{x_{max}}_{x_{1min}} \int^{x_{max}}_{x_{2min}}
F(x_1) F(x_2) \sigma(x_1x_2 s)\theta (1-{4m_h^2\over x_1x_2 s}) dx_1 dx_2,
\end{eqnarray}
with $x_{1min,2min} = 4m_h^2/(x_{max} s)$. Here $\sqrt{s}$ is the total energy of the
$e^\pm$ beams.

The results for the same cases discussed for monochromatic photon
beams are shown in Table 2. The cross section $\sigma(e^+e^-)$ is
smaller than $\sigma(s)$ due to the replacement of $s$ to $x_1x_2
s$. However, we still find that the cross section can be much
larger than that predicted by SM. In SM for $\sqrt{s} = 0.5 $TeV
and $1$ TeV become smaller ($< 0.2$ fb) compared with the case for
monochromatic photon beam. With effects from extra dimensions, for
$\sqrt{s} = 0.5$ TeV, the cross section $\sigma(e^+e^-)$ can still
be as large as 59 (3) fb for $m_h=100$ GeV and 9.4(0.48) fb for
$m_h = 150$ GeV with $M_S = 1.0 (1.5)$ TeV. For $\sqrt{s} = 1$
TeV, the cross sections are much larger. $\sigma(e^+e^-)$ can be
as large as 60 (2.7) fb for $m_h$ as large as 350 GeV with $M_S =
1.01(1.5)$ TeV. For smaller $m_h$, the cross sections are even
larger. These cross sections are much larger than SM predictions.
Observations of Higgs pair events at photon colliders at a level
larger than 1 fb would be indications of effects from extra
dimensions. If the effects due to extra dimensions indeed exist,
production of Higgs pair at photon colliders may even become
practically possible.
Pair production
of Higgs bosons at photon colliders can provide important
information about effects from extra dimensions.

We remark that the angular distribution can also provide important
information about the mechanism for Higgs pair production at
photon colliders. For monochromatic photon colliders the angular
distribution is simply given by $d\sigma/(\sigma d\cos\theta) = 15
\sin^4\theta/16$ which peaks at $\theta=\pi/2$. For back scattered
photon beams, the angular distribution
$d\sigma(e^+e^-)/(\sigma(e^+e^-) d \cos \theta)$ is more
complicated. Here $\theta$ is the angle between the $e^\pm$ beam
direction and the momentum for one of the Higgs boson in the final
state. The distribution also peaks at $\theta = \pi/2$ which is
different than effects due to SM and other new physics, such
anomalous triple Higgs boson coupling\cite{10}. Also the formula
obtained in Eq. (\ref{gg}) can be used for $gg \to HH$ with a
simple replacement of photons to gluons. Effects of extra
dimensions on Higgs pair production can also be studied at hadron
colliders.

To conclude, we have shown that effects due to extra dimensions can induce tree level
$\gamma\gamma \to HH$ scattering. 
Within the allowed range for the scale $M_S$ the
predicted cross sections
can be much larger than those predicted by the Standard Model where $\gamma \gamma \to HH$
only occur at loop level.
New physics due to extra dimensions can be tested using Higgs pair production at
photon colliders.
If the effects due to extra dimensions exist, production of
Higgs pairs at photon colliders may even become practically possible.

Note added. After we have finished this work, we became to aware the
work by Rizzo in Ref.\cite{rrizo} where among other things similar 
calculation was also done for $\gamma \gamma \to HH$. Our formula for
this process agree.

This work is supported in part by the National Science Council of R.O.C under
Grant NSC 88-2112-M-002-041.

\begin{table}
\caption{The cross section $\sigma(s)$ for $\gamma \gamma \to HH$.
a) $\sigma(s)$ (in unit fb) for $\sqrt{s} =0.5$ TeV with $M_S=1.00$ (1.50) TeV.
b) $\sigma(s)$ (in unit pb) for $\sqrt{s} =1.0$ TeV with $M_S = 1.01$ (1.5) TeV.}
\begin{center}
\begin{tabular}{|l|l|l|l|l|l|l|}
\hline $m_h$ (GeV)& n=2&n=3&n=4&n=5&n=6&n=7\\ \hline
a)(fb)&&&&&&\\ \hline
100&2280(114)&941(34)&461(13)&249(6.09)&145(3.273)&91(1.98)\\
150&1155(58)&477(17)&233(6.66)&127(3.09)&74(1.66)&46(1.00)\\
200&274(14)&113(4.08)&55(1.58)&30(0.73)&18(0.39)&11(0.24)\\ \hline
b)(pb)&&&&&&\\ \hline 100&427(7.12)&344(3.76)&297(2.27)&
266(1.46)&243(9.78)&225(6.78)\\ 200&306(5.10)&246(2.70)&213(1.62)&
190(1.04)&174(7.00)&161(4.85)\\ 350&88(1.46)&71(0.77)&61(0.47)&
55(0.30)&50(0.20)&46(0.14)\\
450&7.4(0.124)&6.0(0.066)&5.2(0.039)&4.6(0.025)&4.2(0.017)&3.9(0.012)
\end{tabular}
\end{center}
\end{table}

\begin{table}
\caption{The cross section $\sigma(e^+e^-)$ for $e^+e^-\to \gamma \gamma \to HH$.
a) $\sigma(e^+e^-)$ (in unit fb) for $\sqrt{s} =0.5$ TeV with $M_S=1.00$ (1.50)  TeV.
b) $\sigma(e^+e^-)$ (in unit fb) for $\sqrt{s} =1.0$ TeV with $M_S = 1.01$ (1.5) TeV.}
\begin{center}
\begin{tabular}{|l|l|l|l|l|l|l|}
\hline
$m_h$ (GeV)& n=2&n=3&n=4&n=5&n=6&n=7\\
\hline
a)(fb)&&&&&&\\
\hline
100&59(3.03)&19(0.71)&7.6(0.24)&3.6(0.10)&1.98(0.06)&1.20(0.03)\\
150&9.4(0.48)&3.1(0.12)&1.3(0.04)&0.63(0.02)&0.34(0.009)&0.21(0.006)\\
\hline
b)(fb)&&&&&&\\
\hline
100&5615(271)&3094(106)&1954(50)&
1322(27)&936(15)&683(9.5)\\
200&2607(124)&1481(50)&956(24)&
657(13)&470(7.5)&347(4.7)\\
350&60(2.72)&37(1.17)&25(0.59)&
18(0.33)&14(0.20)&10(0.12)
\end{tabular}
\end{center}
\end{table}

\begin{figure}[htb]
\centerline{ \DESepsf(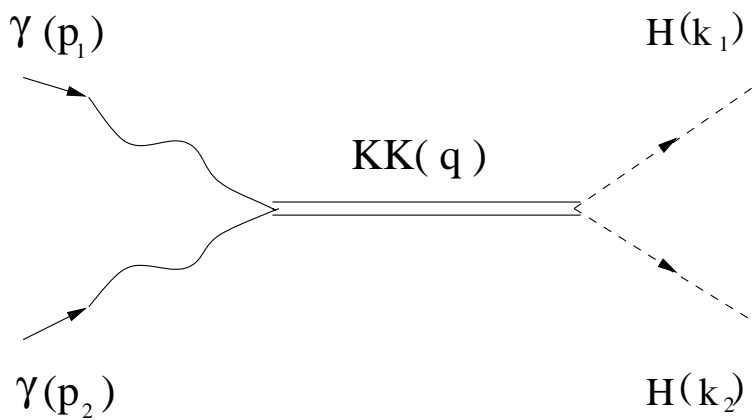 width 10 cm) }
\smallskip
\caption {Tree diagram for KK states contribution to
$\gamma(p_1) \gamma(p_2) \to H(k_1) H(k_2)$.}
\label{penguin}
\end{figure}

\end{document}